\documentclass[aps,prl,twocolumn,superscriptaddress]{revtex4}
\usepackage{graphics}
\usepackage{graphicx}

\begin{document}


\title{E-pile model of self-organized criticality}


\author{A. V. Milovanov}
\email[]{Alexander.Milovanov@phys.uit.no}
\affiliation{Department of Physics and Technology, University of Troms\o, N-9037 Troms\o, Norway}

\author{K. Rypdal}

\affiliation{Department of Physics and Technology, University of Troms\o, N-9037 Troms\o, Norway}

\author{J. J. Rasmussen}
\affiliation{Optics and Plasma Research Department, Technical University of Denmark, Ris\o\,National Laboratory, OPL-129, DK-4000 Roskilde, Denmark}



\begin{abstract}
The concept of percolation is combined with a self-consistent treatment of the interaction between the dynamics on a lattice and the external drive. Such a treatment can provide a mechanism by which the system evolves to criticality without fine tuning, thus offering a route to self-organized criticality (SOC) which in many cases is more natural than the weak random drive combined with boundary loss/dissipation as used in standard sand-pile formulations. We introduce a new metaphor, the e-pile model, and a formalism for electric conduction in random media to compute critical exponents for such a system. Variations of the model apply to a number of other physical problems, such as electric plasma discharges, dielectric relaxation, and the dynamics of the Earth's magnetotail. 
\end{abstract}


\maketitle

A long-standing problem in the study of dissipative non-equilibrium dynamical systems is to explain how long-range temporal correlations with a ``$1/f$" power spectrum develop from local interactions as mass, entropy, and energy are exchanged with the environment. One approach to the problem can be found within a class of discrete lattice models which feature so-called self-organized criticality (SOC). The notion of SOC \cite{SOC} has emerged from the scale-free statistics generated from numerical simulations of the dynamics of such model systems in response to a slow drive. A particularly clear example of SOC is the sand-pile model \cite{SOC}, of which a few variants are known \cite{Tang,Zhang,Dhar,Pietronero}. In sand-piles a small input perturbation triggers chain reactions of redistribution of a certain quantity (like mass or energy) as dictated by a prescribed thresholding condition such as a limit on occupancy per site. In response to the slow drive the system self-adjusts and evolves into a ``critical" state without fine tuning of any external or control parameters. This idea of natural evolution to criticality via a self-organization process is central to all SOC models. The SOC state features no intrinsic length or time scale, its fluctuations are scale-free and are characterized by an inverse power-law power spectral density (PSD). By analogy with the traditional critical phenomena it has been argued that there is universal behavior, which could be described by a set of critical exponents and scaling relations. Crucial to SOC are the microscopic dynamical rules which determine how the perturbations propagate across the system, and which define the universality class of the model \cite{Kadanoff}. Several important ingredients and physical implications of SOC dynamics can be found in e.g., Refs. \cite{Maya,MF,Jensen}. 

\begin{figure}
\includegraphics[width=0.5\textwidth]{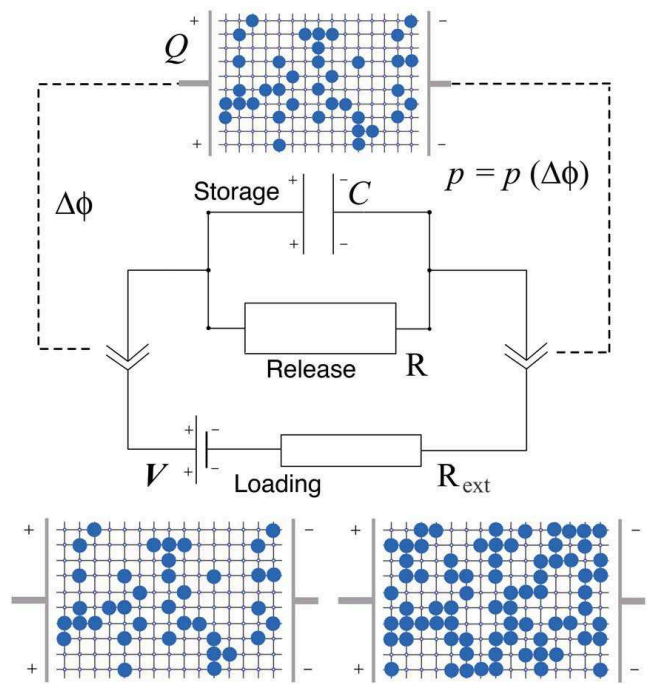}
\caption{\label{} The e-pile model: a dielectric formulation of the SOC properties and of the loading-storage-release cycle, generic of the slowly driven, dissipative, thresholded dynamical systems. Bottom: the changing condition of the lattice for the increasing bias, $\Delta\phi$. Filled circles show nodes in the conducting state. (This figure is in color only in the electronic version.)}
\end{figure}

In this Letter we propose a self-consistent approach to SOC dynamics by including a feedback relation of the system at criticality on the drive. Such an approach  provides a clear mechanism by which the system automatically evolves to criticality without fine tuning any of its parameters, and offers a route to SOC which in many physical settings is more natural than the weak random drive combined with the boundary loss/dissipation as used in standard sand-pile formulations. We analyze the microscopic properties of the SOC state as a transport problem of electric charge. This approach differs from the conventional SOC models, in which one deals with the transport problem of charge-neutral particles whose dynamics is not influenced by an external field. We cast this charge-transport problem into a percolation problem, then combine the concept of percolation with a self-consistent treatment of the interaction between the dynamics on the lattice and the external drive. The SOC literature is rich in metaphors (sand-pile models, forest-fire models, slider-block models, etc.). This is by some interpreted as a lack of clear concepts and rigor. We believe, however, that these metaphors are useful tools when a cross-diciplinary field in science is under conceptual development. In this tradition we will introduce the e-pile model (``e'' stands for ``electric'') as a metaphor for a model SOC system which combines percolation and self-consistent interaction with the external drive.

In the e-pile model the dynamical rules of transport are dictated by the dielectric properties of the medium, and thus make more direct contact with the basic laws of particle motion and the polarization and conductivity response of the system. The model enables one to obtain the critical exponents from known properties of percolation \cite{Stauffer} and of transport of charge in random/disordered media \cite{Gefen,Bouchaud,Dyre,PRB01,PRB07}. The critical exponents are expressible in terms of the percolation indices and are numerically very close to values reported in Refs. \cite{Tang,Zhang}, although those works are based on different assumptions of the nature of the critical state and of the details of the interactions. We take this conformity as a manifestation of the inherent universality of the SOC state.    

{\it Description of the model.} $-$ The model explores the properties of the transition from an insulating to a conducting state in a self-adjusting, disordered dielectric medium. Prospective applications of the model pertain to a variety of physical problems, such as electric plasma discharges, dielectric relaxation, and the dynamics of the Earth's magnetotail \cite{JGR}, but also by mathematical analogy in problems beyond electromagnetics. The SOC regime is achieved by locating the dielectric between the plates of a capacitor and applying an electromotive force as shown in Fig.\ 1. It is assumed that the dielectric consists of a homogenous $d$-dimensional lattice of $N\simeq L^d$ nodes, where $L$ is a characteristic system size (i.e., the distance between the plates). Each node can be in one of two states: (i) a {\it ground} state, in which it acts as an insulator; or (ii) an {\it excited} state, in which it acts as a conductor. The probability for a given node to be in the conducting state depends on the potential difference $\Delta \phi$ across the capacitor and is given by a monotonically growing function $p (\Delta\phi)$ with $p(0) = 0$. Charge can only be conducted between nearest-neighbor nodes and only when these nodes are in the conducting state.

The lattice as a whole acts as an insulator as long as there is no connected cluster of conducting nodes providing at least one conduction path for electric charges across the lattice from one capacitor plate to the other. However, if the electromotive force $V$ exceeds the critical value $p(\Delta \phi _c)=p_c$, where $p_c$ is the percolation threshold probability, the lattice spontaneously develops a conducting path and will as a whole suddenly become conducting, so that its resistance $R$ drops from infinity to a finite value. As  percolation is established, the capacitor starts to discharge, and we shall assume that the discharge time is short compared to the relaxation  time of the external circuit, i.e., $CR\ll CR_{\rm ext}$ (see Fig.\ 1). When the potential difference $\Delta \phi$ drops below $\Delta \phi_c$ the number of  conducting nodes is reduced according to the probability $p (\Delta\phi)$. The lattice then switches to the insulating state, and the cycle repeats. The final state of the system will be ever persisting charging-discharging fluctuations near the percolation point. These fluctuations are self-organized. The work presented in this Letter is based on the conjecture that these self-organized fluctuations are critical, i.e., scale-free. 

This conjecture can be verified by building a quantitative model which can be simulated on a computer. For this we need two additional ingredients: time will have to be discretized and we have to devise a model for how conducting nodes are randomly added or removed from the lattice as the probability $p (\Delta\phi)$ changes from one time step to the next. One conceivable model could be to select $Np(\Delta \phi)$ conducting nodes at random at every time-step. This, however, is hardly what we want from a SOC model, since the lattice will have no memory of its state at the previous time step. A more interesting model would be to add or remove $N \Delta p$ conducting nodes at random every time step. Here $\Delta p$ is the change in occupation probability from the running time step to the next. In this model the lattice remembers its present state when it moves to the next. 

\begin{figure}
\includegraphics[width=0.5\textwidth]{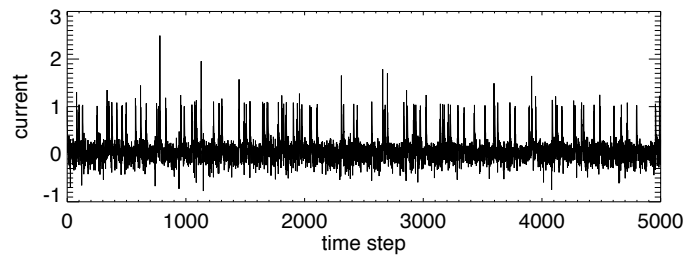}
\caption{\label{} Dynamics of the critical state: cross-lattice electric current in a model in which the conducting network is fully reset after each discharge event. Adapted from Ref. \cite{EPS}.}
\end{figure}

Another ingredient we need is a model for the conductivity across the lattice when it is in the conducting state. One approach to the problem is to assign a resistance and/or complex impedance to the connection between conducting nodes, and use Kirchhoff's equations. By applying  Kirchhoff's equations to the random resistor network one obtains its average frequency- and wave-vector-dependent conductivity, and then the dependence of the conductivity on the order parameter \cite{Stephen}. It is possible to realize a numerical model in which the total resistance over the lattice is calculated from Kirchhoff's rules and graph theory for the running distribution of conducting nodes. Results of this model in the Markovian (memoryless) limit when the entire lattice of conducting nodes is completely updated at random after each discharge event are summarized in Ref. \cite{EPS}. In Fig.\ 2 we reproduce a plot from this study, showing the self-organized fluctuations of the electric current as the voltage drop $\Delta \phi$ fluctuates around the percolation threshold. 

Alternatively, one may consider to inject test-particles and allow them to hop between the nearest-neighbor conducting states in the direction of the voltage drop. This setting has important features in common  with broadly studied directed-percolation and forest-fire models (Ref. \cite{MF} for a brief overview). The analogy is particularly clear if one distinguishes three types of nodes: (i) {\it empty} nodes, i.e., nodes in the insulating or ground state; (ii) {\it green trees}, i.e., nodes in the conducting state, which for the time being do not contain particles; and (iii) {\it burning trees}, i.e., nodes in the conducting state which contain at least one particle. The charged particles are thought of as fire in this model. The difference is that the green trees are not burned by the fire, i.e., they remain green after having transmitted the charge, and their population is only affected by the environmental change in terms of the changing bias, $\Delta\phi$. There are variants of the model in which not more than one (fermion-type) or arbitrary many (boson-type) particles per node are allowed. 

The purpose of this Letter is, however, not to present a numerical study to prove the existence of critical behavior, but rather to devise analytical methods to compute the critical exponents, assuming that the fluctuations are critical. Our analytical results conform with previously reported results of numerical simulation of sand-pile models \cite{Tang,Zhang} and with known estimates in the mean-field limit \cite{MF}.    

{\it The percolation exponents.} $-$ Here for completeness of the discussion we briefly review the definitions of the percolation exponents $\nu$, $\beta$, and $\mu$ (Refs. \cite{Stauffer,Gefen}). In the vicinity of the percolation point $p_c$, the percolation correlation (the pair connectedness) length diverges as $\xi \propto |p - p_c| ^{-\nu}$ for both $p > p_c$ and $p < p_c$. For $p > p_c$, the probability to belong to an infinite cluster is $P_{\infty} \propto (p - p_c)^{\beta}$ and the dc conductivity of the cluster is $\sigma _{\rm dc} \propto (p - p_c) ^{\mu}$. The infinite cluster is a  fractal, with the Hausdorff dimension $d_f = d - \beta / \nu$.        

{\it ac-conduction exponent.} $-$ If a fractal conducting network is exposed to an ac field, its frequency-dependent conductivity  takes the form $\sigma (\omega) \propto \omega ^{\eta}$ where $\omega$ is the applied frequency. The ac-conduction exponent $\eta$ has been obtained by Gefen {\it et al.} \cite{Gefen} from a random-walk model of particle diffusion on percolation clusters: $\eta = \mu / (2\nu + \mu - \beta)$. Using known estimates \cite{Stauffer} for $\nu$, $\beta$, and $\mu$ it is found that $\eta \simeq 0$,~0.3, and~0.6 for $d=1$,~2, and~3, respectively. The mean-field result, holding for $d \geq 6$, is $\eta = 1$. (Note that by mean-field we mean mean-field percolation, which is not to be confused with the mean-field theories of SOC \cite{MF}. Despite of some conformity in the critical exponents there are subtleties of definition of the upper critical dimension: $d_u=6$ for percolation, and $d_u=4$ for SOC.) 

{\it Power spectral density exponent.} $-$ This exponent is obtained from linear response theory. Let ${\bf E} (t, {\bf r})$ be the electric field at time $t$ at point ${\bf r}$ in the bulk of the dielectric. The polarization response to this field is defined as ${\bf P} (t, {\bf {r}}) = \int _{-\infty}^{+\infty} \chi (t - t^{\prime}) {\bf E} (t^{\prime}, {\bf {r}}) dt^{\prime}$ where $\chi (t - t^{\prime})$ is a response function. In the frequency domain, ${\bf P} (\omega, {\bf {r}}) = \chi (\omega) {\bf E} (\omega, {\bf {r}})$ where $\chi (\omega)$ is the frequency-dependent complex susceptibility. The power spectral density (PSD) of the polarization field is given by $S(\omega) = \langle |{\bf P}(\omega, {\bf {r}})|^2\rangle = |\chi(\omega)|^2\langle |{\bf E}(\omega, {\bf {r}})|^2\rangle$ where the angle brackets denote an ensemble average. One can see that $|\chi(\omega)|^2$ is the PSD of the polarization field when the driving electric field is an uncorrelated white noise signal. With use of the Kramers-Kronig relation $\chi(\omega) \propto \mathrm{P} \int d\omega^{\prime} \sigma (\omega^{\prime}) / \omega^{\prime} (\omega^{\prime} - \omega)$ it is found that $\chi (\omega) \propto \omega ^{\eta - 1}$. Hence, $S(\omega) \propto 1/\omega ^\alpha$ with $\alpha = 2-2\eta$. We have $\alpha \simeq 2$,~1.4, and~0.8 for $d=1$,~2, and~3. The mean-field result is $\alpha = 0$. Note that, in 1 dimension the PSD of the SOC state conforms with the PSD of a Brownian random-walk process, i.e., $\alpha = 2$. The result in 2 dimensions, i.e., $\alpha \simeq$~1.4, may be used to explain the low-frequency fluctuation spectrum in the current sheet of the Earth's magnetotail, which for years have been an issue in geo-space plasma research \cite{JGR}. 

{\it Dispersion-relation exponent.} $-$ The dispersion relation shows how the lifetime of an ``activation" cluster scales with its size \cite{Zhang}. In the e-pile model, by activation cluster one would mean a connected cluster of conducting nodes with particles. The finite lifetime of such clusters is due to diffusion of the particles in the conducting domain. We assume that the diffusion occurs as a result of hopping of charged particles between the nearest-neighbor conducting nodes and that there is a characteristic hopping time $\Delta t$ after which a particle changes its site of residence by hopping at random from the hosting node to one of the nearest-neighbor conducting nodes. Given an activation cluster of linear size $\ell$ one estimates its lifetime to be $t \sim \ell\Delta t$, where we have set the microscopic lattice distance to unity for simplicity. The hopping time is estimated as the microscopic conduction time, i.e., the inverse of the average size-dependent conductivity of the cluster, $\sigma (\ell)$. The latter scales with the cluster size as $\sigma (\ell) \propto \ell ^{-\eta}$ \cite{Gefen} leading to $t\sim \ell / \sigma (\ell) \propto \ell ^{1+\eta}$. Hence, the dispersion exponent is $z = 1+\eta$. It is found that $z \simeq$~1.3 and~1.6 for $d=2$ and~3, respectively. These values are numerically very close to the values obtained in the Zhang model \cite{Zhang} but with different analytical expression. Similar values have also been found in the numerical simulation of Ref. \cite{Tang}. In the mean-field limit we have $z = 2$. This result conforms with the results of Ref. \cite{MF}, in which a systematic mean-field treatment of SOC properties is given. In 1 dimension we have $z=1$.               

{\it Relaxation-function exponent.} $-$ The issue of the dispersion relation is closely connected with the issue of relaxation of charge-density inhomogeneities on a fractal lattice at percolation. Here we propose a self-consistent approach to the problem, in which the dynamics of decay of the activation clusters is governed by the electric field fluctuations, produced by the electric charges themselves. Let $\hat\rho (t, {\bf r})$ be the charge density at time $t$ at point ${\bf r}$ in the bulk of the dielectric. These charges inside the dielectric generate the electric field inhomogeneity $\hat {\bf E} (t, {\bf r})$ whose divergence is $\nabla\cdot \hat {\bf E} (t, {\bf r}) = 4\pi\hat\rho (t, {\bf r})$. The polarization response to this field is $\hat {\bf P} (t, {\bf {r}}) = \int _{-\infty}^{+\infty} \chi (t - t^{\prime}) \hat{\bf E} (t^{\prime}, {\bf {r}}) dt^{\prime}$ and its divergence is $\nabla\cdot\hat {\bf P} (t, {\bf r}) = -\hat\rho (t, {\bf r})$. The flow function is defined by $\hat {\bf j} (t, {\bf r}) = \partial_t \hat {\bf P} (t, {\bf r})$ where $\partial_t$ denotes time derivative. From the equation of continuity we find $\partial_t \hat\rho (t, {\bf r}) = - 4\pi \int _{-\infty}^{+\infty} \sigma (t - t^{\prime}) \hat\rho (t^{\prime}, {\bf {r}}) dt^{\prime}$ where we have introduced $\sigma (t - t^{\prime}) = \partial_t \chi (t - t^{\prime})$. If the dynamics start at time $t=0$, then due to causality  $\int _{-\infty}^{+\infty} dt^{\prime}=\int _{0}^{t} dt^{\prime}$. One readily rewrites the dynamic equations in the Laplace domain to find $s\hat\rho (s, {\bf r}) + 4\pi \sigma (s) \hat\rho (s, {\bf {r}}) = \hat\rho (0, {\bf r})$  \cite{PRB07}. Here $\hat\rho (0, {\bf r})$ is the initial charge density, which is set to 1 for simplicity. Assuming for the conductivity $\sigma (s) = \kappa s^\eta$ with $\kappa$ a constant, then setting $\eta = 1-\zeta$ we have $\hat\rho (s, {\bf r}) = 1 /(s+4\pi\kappa s^{1-\zeta})$. The inverse Laplace transform of this is the Mittag-Leffler function \cite{Klafter} whose initial-time behavior is a stretched-exponential decay function $\hat\rho (t, {\bf r}) \propto \exp \left[-4\pi\kappa t^\zeta / \Gamma (\zeta + 1)\right]$. This stretched-exponential behavior was suggested for SOC systems in Ref. \cite{Tang}, without a rigorous derivation. For the relaxation-function exponent $\zeta = 1-\eta$ we have $\zeta \simeq$~0.7 for $d=2$ and $\zeta \simeq~0.4$ for $d=3$. The mean-field value is $\zeta = 0$. Note that 1-dimensional critical relaxation is exponential: $\zeta=1$. 

It has been shown that the properties of ac conduction and dielectric relaxation in disordered dielectrics are expressible in terms of fractional-derivative equations, and a systematic derivation of the fractional relaxation and fractional diffusion equations was presented in Ref.\  \cite{PRB07}. These results suggest that systems in states of self-organized criticality are described by fractional kinetics. 

{\it Size-distribution exponent.} $-$ In sand-pile models the exponent $\tau$ is introduced to characterize the power-law distribution of avalanche sizes measured as  the number of sites involved in the dynamics of isolated relaxation events \cite{SOC,Tang,Zhang}. These notions are naturally present  in  e-piles, and the $\tau$ values are readily obtained from Eq.\ (5) of Ref. \cite{Tang} where one replaces the ``noise" exponent $\phi$ with $\alpha = 2(1-\eta)$, the fractal dimension $D$ with $d_f$, and the dynamical exponent $z$ with $1+\eta$ to find $d_f (3-\tau) = \alpha (1+\eta)$. The final result depends on the convention of {\it order} of limits. We have  to specify which limit to be taken first: the thermodynamic limit $L\rightarrow\infty$, or the zero-frequency limit $\omega\rightarrow 0$. If the thermodynamic limit is taken first, we make use of the percolation dimension $d_f = d - \beta / \nu$ and  then apply $\eta = \mu / (2\nu + \mu - \beta)$ to obtain $\tau\simeq$~2.0 for $d=2$ and $\tau\simeq~2.5$ for $d=3$. The mean-field result deriving from $\eta = 1$ and $\alpha = 0$ is $\tau=3$. If the zero-frequency limit is taken first, then one has to replace $d_f$ with $d$, then to set $\eta = 0$ in the dc (frequency-independent) limit to get $z=1$, $\alpha = 2$, and $\tau = 3-2/d$. This dependence of $\tau$ on $d$ was reported in Ref.\ \cite{Zhang}. Numerically, $\tau\simeq$~2.0 for $d=2$ and $\tau\simeq~2.3$ for $d=3$. The corresponding mean-field value is obtained for $d_f = 4$ yielding $\tau = 2.5$, in agreement with Ref.\ \cite{MF}. Note that the definitions of $\tau$ in Ref.\ \cite{Tang} and in Refs.\ \cite{Zhang,MF} differ by 1. 

{\it Summary.} $-$ In conclusion, we have introduced a new self-consistent approach to realize SOC dynamics: the e-pile. We have provided analytical methods to calculate the main critical exponents characterizing the dynamics, under the assumption that the fluctuations are critical. Other critical exponents may be obtained if desired. The results agree with previously reported results from numerical investigations of the sand-pile models.  Numerical simulation of the e-pile dynamics is under way for comparison with the analytical predictions.  


\begin{acknowledgments} 
Discussions with our colleagues A. V. Chechkin, G. Consolini, M. P. Freeman, B. V. Kozelov, R. Sanchez, A. S. Sharma, and F. Zonca are gratefully acknowledged. This work was funded under the project No 171076/V30 of the Norwegian Research Council.
\end{acknowledgments}


\end{document}